\documentclass[%
superscriptaddress,
amsmath,amssymb,
aps,
prb,
floatfix,
twocolumn
]{revtex4-2}

\usepackage{graphicx}% Include figure files
\usepackage{dcolumn}% Align table columns on decimal point
\usepackage{bm}% bold math
\usepackage{hyperref}% add hypertext capabilities
\hypersetup{colorlinks=true,citecolor={blue},linkcolor={blue},urlcolor={blue}}
\usepackage{physics}
\usepackage{xcolor}
\usepackage{bbold}

\newcommand{\pcsadd}{Center for Theoretical Physics of Complex Systems, Institute for Basic Science(IBS), Daejeon, Korea, 34126}
\newcommand{\ustadd}{Basic Science Program, Korea University of Science and Technology (UST), Daejeon 34113, Republic of Korea}

\newcommand{\tlam}{\tilde{\lambda}}

\begin{document}

\title{Enhancement of Superconductivity in the Fibonacci Chain}
% Force line breaks with \\
% \thanks{A footnote to the article title}%

\author{Meng Sun}
    \affiliation{Faculty of Science, Beijing University of Technology, Beijing, China, 100124}
    \affiliation{\pcsadd}
    \email{msun\_89@bjut.edu.cn}

\author{Tilen \v{C}ade\v{z}}
    \affiliation{\pcsadd}

\author{Igor Yurkevich}
    \affiliation{School of Computer Science and Digital Technologies, Aston University, B4 7ET Birmingham, United Kingdom}
    \affiliation{\pcsadd}
    
\author{Alexei Andreanov}
    \affiliation{\pcsadd}
    \affiliation{\ustadd}
    \email{aalexei@ibs.re.kr}

\date{\today}% It is always \today, today,
%  but any date may be explicitly specified

\begin{abstract}
    We study the interplay between quasi-periodic disorder and superconductivity in a 1D tight-binding model with the quasi-periodic modulation of on-site energies that follow the Fibonacci rule and all the eigenstates are multifractal. 
    As a signature of multifractality, we observe the power-law dependence of the correlation between different single-particle eigenstates as a function of their energy difference. 
    By computing numerically the superconducting transition temperature, we find the distribution of critical temperatures, analyze their statistics and estimate the mean value and variance of critical temperatures for various regimes of the attractive coupling strength and quasi-periodic disorder. 
    %We find an enhancement of the critical temperature as compared to the analytical results that are based on strong but hardly justifiable assumption of self-averaging nature of multiple characteristics of the system. 
    We find an enhancement of the critical temperature as compared to the analytical results that are based on strong assumptions of absence of correlations and self-averaging of multiple characteristics of the system, which are not justified for the Fibonacci chain.
    For the very weak coupling regime, we observe a crossover where the self-averaging of the critical temperature breaks down completely and a strong sample-to-sample fluctuations emerge.
\end{abstract}

\maketitle

\section{Introduction}

Fractals are intricate geometric objects that are self-similar across different scales~\cite{ben-avraham2000diffusion,mandelbrot2020fractals}. 
The concept of fractality has revolutionized the development of novel materials and devices, offering unique properties and applications. 
Materials featuring fractal structures showcase exceptional characteristics typically not observed in non-fractal counterparts. 
One remarkable example is the recent advancement in fractal graphene-based materials~\cite{zhao2019disorder}. 
These materials display remarkable mechanical strength, electrical conductivity, and thermal stability, making them highly suitable for a diverse range of applications, including energy storage and sensing.
The fractal structures have proven to be efficient in photovoltaic devices~\cite{yeon2020fractal} since they enhance light absorption and significantly improve the efficiency of solar cells. 
For instance, the utilization of fractal-shaped nanowires in solar cells has led to heightened light trapping and absorption compared to conventional designs~\cite{fazio2016strongly}. 

From a theoretical perspective, there have been notable efforts to explore the conditions under which fractal geometry can enhance a  property critical for applications, such as superconductivity. 
Following the development of the microscopic theory of superconductivity by Bardeen, Cooper, and Schrieffer (BCS)~\cite{bardeen1957microscopic,bardeen1957theory,allen1983theory}, the influence of disorder on superconductivity garnered considerable attention~\cite{anderson1959theory,strongin1970destruction,imry1981destruction,yurkevich2001nonlinear}. 
Early studies~\cite{falko1995multifractality,feigelman2007eigenfunction,feigelman2010fractal,mayoh2015global} suggested that a superconducting phase could emerge when the Fermi energy (\(E_F\)) resides in the region of the Anderson mobility edge due to strong correlations between fractal wavefunctions. 
Subsequent research predicted an increase in critical temperature even in quasi-one-dimensional (1D) wires~\cite{lowe2021disorder, kagalovsky2021disorder}, quasi-2D materials~\cite{mayoh2015global}, and weakly disordered two-dimensional (2D) systems~\cite{fyodorov1997strong}.

While many studies (see Ref.~\onlinecite{kravtsov2010superconducting} and references there) focused on situation when transition in a clean system described by the standard BCS-type mean-field theory is modified by disorder inducinng significant overlap between multifractal wavefunctions with different eigenenergies, there are also quasiperiodic materials that possess these features intrinsically without extrinsic disorder. 
Quasiperiodic systems, readily realized experimentally in various structures like artificial atomic chains and quasi-2D semiconducting heterostructures~\cite{yan2019engineered}, serve as examples. 
The Fibonacci chain~\cite{jagannathan2021the,chiaracane2021quantum,mace2019many,mace2016fractal,tanese2014fractal}, a one-dimensional quasiperiodic structure closely related to three-dimensional icosahedral quasicrystals~\cite{fang2015an}, offers an intriguing realm for superconductivity studies. 
The energy spectrum in this system exhibits a Cantor set-type fractal structure~\cite{kohmoto1987critical}, and the multifractal eigenfunctions demonstrate long-range power-law spatial and temporal correlations.
It is therefore a natural testbed for the effect of fractality on the superconducting properties.

Phenomenological arguments were put forward~\cite{kravtsov2012wonderful} suggesting multifractal correlations of wavefunctions enhance superconductivity.
This was later tested close to the Anderson transition within a mean-field approximation~\cite{feigelman2010fractal}.
The difficulty is that now one has to solve a disordered gap equation, without the simplifications brought in by translation invariance.
A common approach is to average the gap equation and ignore the correlations~\cite{mayoh2015global}.
As we demonstrate in this work, neglecting the correlations removes an important enhancement of the critical temperature.

The outline of the paper is as follows.
We define the model and study its spectral correlation function in Sec.~\ref{sec:model}.
Then the mean-field approximation to superconductivity in the model and the behavior of the average critical temperature are studied in Sec.~\ref{sec:mf}.
The breakdown of self-averaging of the critical temperature and the crossover in the coupling strength are discussed in Sec.~\ref{sec:self-av}.
This is followed by Conclusions.

\section{Model \& Spectral correlation function}
\label{sec:model}

We consider the 1D Fibonacci chain, which serves as a fundamental model representing quasicrystals. 
This chain exhibits several noteworthy properties, as outlined in a recent study by Jagannathan et al.~\cite{jagannathan2021the}: 
(i) deterministic construction - the Fibonacci chain is constructed following a well-defined deterministic algorithm; 
(ii) finite number of possible configurations - despite its complexity, the Fibonacci chain possesses a finite number of possible configurations allowing detailed analysis; 
(iii) multi-fractal eigenstates - one of the remarkable features of the Fibonacci chain is that its eigenstates exhibit multifractal behavior for all values of the onsite potential \(h\) leading to intricate patterns with varying degrees of complexity and self-similarity, regardless of the specific values of the onsite potential.

Here we focus on a chain model with on-site energies arranged according to the Fibonacci rule. 
The tight binding Hamiltonian, 
\begin{gather}
    \hat{H}_F = -\sum_i  \left( \hat{c}_i^\dagger \hat{c}_{i+1} + \hat{c}^\dagger_{i+1}\hat{c}_i + h_i \hat{c}_i^\dagger \hat{c}_i\right)\,,
    \label{eq:2-1}
\end{gather}
describes particles hopping between lattice sites with dimensionless (measured in units of hopping amplitude) on-site potential \(h_i\). 
The potential takes two values \(\pm h\) which are arranged according to the Fibonacci sequence rule \(\sigma: A \to AB \,, B \to A\).
The \(n\)th Fibonacci word \(W_n\) is the concatenation of two previous ones \(W_n = \left[ W_{n-1}, W_{n-2} \right]\).
To construct the Fibonacci type potential for a system of size \(L\), we first write down a long enough Fibonacci sequence, then cut a segment containing \(L\) consecutive letters, and make the substitution \(A \to h\) and \(B \to -h\). 
The number of different segments \(N=L/2\) (\(N=(\left(L-1\right)/2\)) for \(L\) even (odd)~\cite{chiaracane2021quantum}.
In this way, we generate ensemble of \(N\) different realizations of on-site energy arrangements, each being a subset of the Fibonacci sequence.

Some properties of the eigenstates of the Fibonacci chain have been studied recently~\cite{niu1990spectral,jagannathan2021the,fan2021enhanced}. 
For example, a perturbative renormalization group analysis was used to analytically determine fractal dimensions for the off-diagonal Fibonacci chain~\cite{mace2016fractal} in the weak potential strength limit (\(h \ll 1\)). 
For the issue of superconducting transition, the most important property of multifractal systems~\cite{feigelman2007eigenfunction,kravtsov2012wonderful}, is the overlap of different eigenstates described by the correlation of two single particle wavefunction~\cite{kravtsov2012wonderful},
\begin{gather}
    \label{eq:2-2}
    C\left( \omega \right) = L^d \sum_{\mathbf{r},n,m} \langle \abs{\psi_n(\mathbf{r})}^2 \abs{\psi_m(\mathbf{r})}^2 \delta\left( \epsilon_m - \epsilon_n -\omega\right)\rangle \,,
\end{gather}
where \(L^d\) is the system volume, \(\psi_n (\mathbf{r})\) and \(\epsilon_n\) are the eigenstate and eigenenergy of the Hamiltonian~\eqref{eq:2-1}, respectively. 
This function demonstrates power-law decay at the Anderson transition~\cite{feigelman2007eigenfunction,cuevas2007two},
\begin{gather}
	\label{eq:2-3}
	C\left(\omega \right) = \left( \frac{E_0}{\abs{\omega}}\right)^\gamma\,,
\end{gather}
in some frequency domain \(\delta_L < \omega <E_0\), where \(\delta_L\) is the mean level spacing and \(E_0\) is the energy scale related to the fractal length. 
The power-law exponent \(\gamma\) is connected to the multifractal dimension by a simple relation~\cite{chalker1990scaling,feigelman2010fractal}: \(\gamma = 1 - \frac{d_2}{d}\).
 
\begin{figure}
    \centering
    \includegraphics[width=0.49\textwidth]{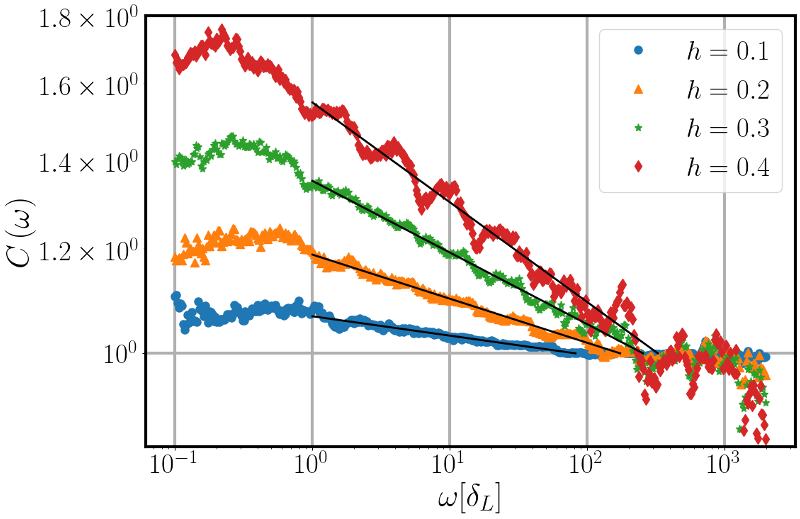}
    \caption{
        The correlation function~\eqref{eq:2-3} of the Fibonacci chain for fixed system size \(L=2000\) and several disorder strengths. 
        The solid black lines are the power-law fits, Eq.~\eqref{eq:2-3}.
        The fitted values of \(\gamma\) and \(E_0\) are \(\gamma \sim 0.0146,~0.0333,~0.0542,~0.0754\) and \(E_0 \sim 82.1,~173.3,~254.6,~325.9 ~\left[\delta_L \right]\) for \(h=0.1,~0.2,~0.3,~0.4\).
    }
    \label{fig:2-1}
\end{figure}

We confirm the power-law decay of the correlation in the Fibonacci chain, see Fig.~\ref{fig:2-1}, for different disorder strengths, by numerical diagonalization of the Hamiltonian~\eqref{eq:2-1} and averaging over different realizations, i.e. different slices of the length \(L\) cut from \(n\)-th Fibonacci word.
Using the numerically computed correlator~\eqref{eq:2-3}, we further estimate the upper energy scale \(E_0\) and the exponent \(\gamma\) from the power-law fits shown in Fig.~\ref{fig:2-1}.
To the best of our knowledge, this correlation function has not been studied yet for the Fibonacci chain.

%   Figure 1

\section{Mean-field superconductivity}
\label{sec:mf}

The spinful fermions on a tight-binding chain with local attraction are described by the negative-\(U\) Hubbard Hamiltonian, 
\begin{align}
    \label{eq:3-0}
    \hat{H} = \sum_{\sigma}\hat{H}_{F,\sigma} +U \, \sum_{i=1}^L {\hat n}_{i\uparrow} {\hat n}_{i\downarrow}
\end{align}
where the single-particle part \(\hat{H}_{F,\sigma} \) is given by Eq.~\eqref{eq:2-1} for each of spin components \(\sigma = \uparrow, \downarrow\). 
The second term, with \({\hat n}_{i\sigma}={\hat c}_{i\sigma}^{\dagger}{\hat c}_{i\sigma}\) being the occupation number operator of electrons with spin \(\sigma\) on \(i\)-th site, 
is the attractive Hubbard interaction with dimensional coupling constant \(U\).

To investigate the superconducting properties we write the Hamiltonian in the single particle eigenbasis of \(\hat{H}_{F,\sigma} \) 
\begin{gather}
    {\hat c}_{i\sigma}=\sum_n\,\psi_n(i)\,{\hat c}_{n\sigma}\,,
\end{gather}
following~\cite{feigelman2007eigenfunction, ma1985localized}, and keep only the terms most relevant for the superconductivity 
\begin{align}
    \label{eq:3-1}
    \hat{H} &= \sum_{n\sigma} \epsilon_n \hat{c}^\dagger_{j\sigma} \hat{c}_{n\sigma} + U\,\sum_{nm} M_{nm} \hat{c}^\dagger_{n\uparrow} \hat{c}_{n\downarrow}^\dagger \hat{c}_{m\uparrow} \hat{c}_{m\downarrow} \\
    \label{eq:3-2}
    M_{nm} &= \sum_i\abs{\psi_n(i)}^2 \abs{\psi_m(i)}^2,
\end{align}
where \(\epsilon_n\) is the single-particle energy of state \(n\); and \(\sigma = \lbrace \downarrow, \uparrow \rbrace\) is the spin label. 
The mean-field approach~\cite{mahan2008many} leads to the gap equation
\begin{gather}
    \label{eq:3-4}
    \Delta_n = \frac{|U|}{2}\sum_m \frac{M_{nm} \Delta_m}{\varepsilon_m} \tanh \left( \frac{\epsilon_m}{2T} \right).
\end{gather}
where \(\varepsilon_n=\sqrt{\epsilon_m^2 + \Delta_m^2}\), and the gap function is defined as an anomalous Green function, 
\(\Delta_n=\langle {\hat c}_{n\uparrow}{\hat c}_{n\downarrow}\rangle\).
The transition is signalled by the appearance of a non-zero \(\Delta_n\).

%( \igor{\textbf{may we refer to more papers?}}
The routine approach to analysing the transition is based on few assumptions~\cite{feigelman2007eigenfunction,feigelman2010fractal,mayoh2015global}:
\begin{enumerate}
    \item density of states and the wavefunctions are uncorrelated; 
    \item density of states \(\nu_0\) is self-averaging and energy-independent in the window of the Debye frequency \(\epsilon_D\)  around the Fermi energy; 
    \item all gap functions \(\Delta_n\) are self-averaging, and, finallly, 
    \item there is no correlation between the wavefunctions overlap integral \(M_{nm}\) and the gaps \(\Delta_n\).
\end{enumerate}
Only under all the mentioned above conditions, the gap equations acquire the following form in the continuous limit after averaging over the realisations:
\begin{gather}
    \label{eq:3-5}
    \Delta(\epsilon) = \frac{\lambda}{2} \int^{\epsilon_D}_{-\epsilon_D} \frac{d \epsilon'}{\varepsilon(\epsilon')} C(\epsilon-\epsilon') \tanh \left( \frac{\varepsilon (\epsilon')}{2T} \right) \,\Delta(\epsilon') \,.
\end{gather}
Here another dimensionless coupling constant is introduced \(\lambda=\nu_0\,|U|\).
Further assuming that all the gaps \(\Delta_n\) vanish at the transition, i.e. in the continuous limit \(\Delta(\epsilon)=0\), and that the Debye frequency is much larger than the fractal scale \(E_0\), leads to the following equation for the critical temperature,
\begin{gather}
    1 = \lambda \int_0^{\epsilon_D} \frac{C\left(\epsilon\right)}{\epsilon} \tanh\left( \frac{ \epsilon}{2T_c^A}\right) d\epsilon\,,
\end{gather}
which admits the solution,
\begin{gather}
    \label{eq:3-6}
    T_c^{A} = \epsilon_D \mathcal{D}\left(\gamma\right) \left[ 1+\frac{\gamma}{\lambda} \left( \frac{\epsilon_D}{E_0} \right)^\gamma \right]^{-\frac{1}{\gamma}}\,,
\end{gather}
with
\begin{gather}
    \label{eq:3-7}
    \mathcal{D}\left(\gamma\right) = \left[ 2\gamma \left(2^{\gamma+1}-1\right) \Gamma\left(-\gamma\right) \zeta\left(-\gamma\right)\right]^{\frac{1}{\gamma}}\,,
\end{gather}
and \(\zeta\left(x\right)\) is the Riemann \(\zeta\) function~\cite{mayoh2015global}.

\begin{figure}
    \centering
    \includegraphics[width=0.48\textwidth]{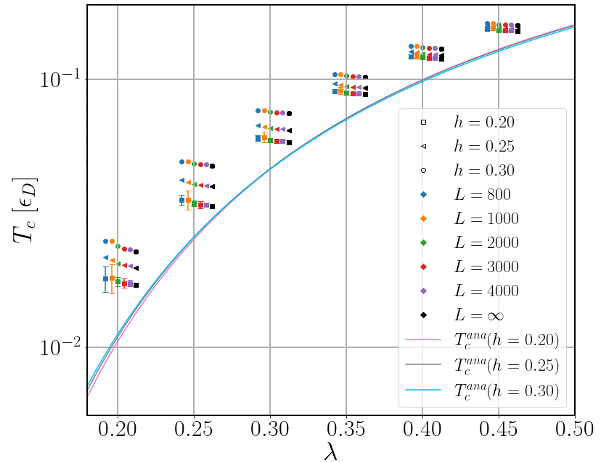}
    \includegraphics[width=0.46\textwidth]{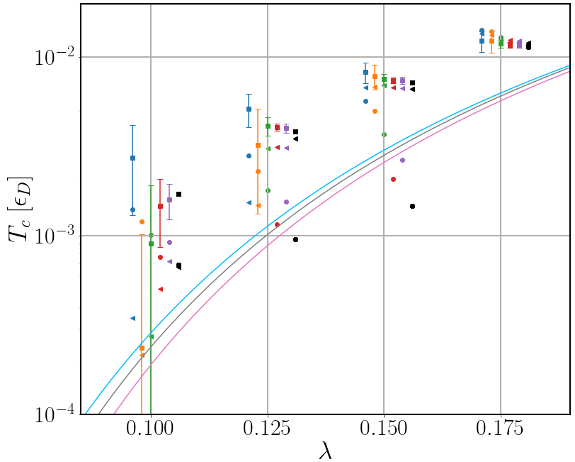}
    \caption{
        Average critical temperature for different system sizes (color of the markers) and different disorder strengths (shape of the markers).
        The black points are the result of the extrapolation to the infinite size.
        The solid lines are the analytical result of Ref.~\onlinecite{mayoh2015global}.
        The vertical bars show the temperature variance for disorder \(h=0.3\).
        For different system sizes and disorder strengths, the markers are shifted to the left and right relative to the green markers.
        Top: larger couplings \(\lambda\).
        Bottom: small couiplings \(\lambda\).
    }
    \label{fig:3-1}
\end{figure}

We extract the values of \(E_0\) and \(\gamma\) from the correlation function~\eqref{eq:2-2} which takes the power-law scaling form~\eqref{eq:2-3}, as we have verified in Sec.~\ref{sec:model}.
%Under assumption that all entries are self-averaging, the analytical result for the critical temperature~\eqref{eq:3-6} implies that the fractal energy scale \( E_0 \) and the exponent \( \gamma \) are found by averaging over realisation of Fibonacci potentials.
%The result is shown in Fig.~\ref{fig:2-1}.
With these averaged parameters, we can evaluate the critical temperature by Eq.~\eqref{eq:3-6}. 
We show in Fig.~\ref{fig:3-1} by the solid line the critical temperature computed via Eq.~\eqref{eq:3-6} as the function of coupling strength for the Fibonacci chain system with different disorder strengths.

However, this approach is based on at least four assumptions outlined above which are hard to justify. 
Instead, we compute the critical temperature numerically without a priory assumptions on statistics and correlations between various entries present in Eq.~\eqref{eq:3-4}. 
We solve the gap equation~\eqref{eq:3-4} in the limit of vanishing gaps \(\Delta_n\),
\begin{gather}
    \label{eq:3-8}
    \Delta_n = \frac{\lambda}{2\nu_0}\sum_m^{\abs{\epsilon_m} < \epsilon_D} \frac{M_{nm} }{\epsilon_m} \tanh \left( \frac{\epsilon_m}{2T_c} \right) \Delta_m.
\end{gather}
to find the critical temperature \(T_c\) numerically for every realization of the Fibonacci potential, and then analyze the statistics of the ensemble of critical temperatures: their distribution function, mean value and variance.

%   Figure 2

The results for the average critical temperature of the Fibonacci chain, computed along the above lines, are presented in Fig.~\ref{fig:3-1} for several system sizes \(L\), disorder strengths \(h\) and couplings \(\lambda\).
%The points show the averaged critical temperature for different disorder strengths (shape of the markers) and lengths (color of the markers) for a given coupling strength \(\lambda\).
For convenience of presentation, the points are manually shifted horizontally for fixed couplings \(\lambda\).
The vertical bars show the standard deviation of the critical temperature.
For convenience we only show the error bars for the case \(h=0.30\) -- the error bars for other disorder strengths show similar behavior.
%in Fig.~\ref{fig:3-1} for clearness.
At last, we estimate the critical temperature in the thermodynamic limit by the finite-size extrapolation.
The results are labelled with black markers.

We observe that over a wide range of couplings the average critical temperature is self-averaging with small variance.
The variance increases significantly as the coupling strength is decreased, as seen in the bottom plot of Fig.~\ref{fig:3-1}.
This suggests the existence of a crossover coupling strength \(\tlam\) below which the critical temperature starts to lose its self-averaging property and sample-to-sample fluctuations become important.
Detailed discussion of this crossover and its properties is provided in the next section.

\begin{figure}
\centering
    \includegraphics[width=0.40\textwidth]{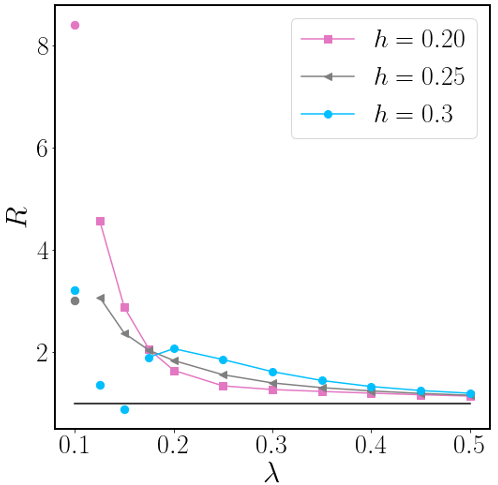}
    \caption{
        The enhancement ratio \(R = T_c^N/T_c^A\) vs coupling \(\lambda\) at \(L=4000\) for different disorder strengths \(h\): \(0.2\) (squares), \(0.25\) (triangles) and \(0.3\) (circles).
        \(T_c^N\) is the critical temperature computed numerically, \(T_c^A\) is the critical temperature predicted by Ref.~\onlinecite{mayoh2015global}. 
        The markers are connected by solid lines in the region \(\lambda > \tlam\).
        The black line corresponds to \(R=1\) and is shown for convenience. 
%        \alexei{AA: what is the black line? and what is \(\tlam\)?
%        \meng{Meng: black line is the reference line fro $R = 1$. $\tlam_c^{max}$ should be $\tlam$.}}
    }
    \label{fig:3-2}
\end{figure}

The main result shown in Fig.~\ref{fig:3-1}, is the clear discrepancy between the two procedures - assuming self-averaging properties and absence of correlation followed by analytic solution of the Eq.~\eqref{eq:3-6}, 
and a straightforward numerical analysis of random critical temperatures found from the exact Eq.~\eqref{eq:3-8} with no assumptions at all.
That is although the critical temperature self-averages, this self-averaging value is different from the solution of Eq.~\eqref{eq:3-6}.
In most regions of the coupling strength, we find an enhancement of the critical temperature compared to the analytical formula, Eq.~\eqref{eq:3-8}.
By denoting the average critical temperature following from Eq.~\eqref{eq:3-8} as \(T_c^N\), we calculate the enhancement ratio \(R  = T_c^N /T_c^A\) as shown in Fig.~\ref{fig:3-2}.
%and in the right panel of the figure we further compare the critical temperature \(T_c^N\) and \(T_c^A\) with the standard BCS result: \(T_c^B\), expected for a clean $h=0$ system.
For convenience we connected by lines the ratios for the coupling strengths above the self-averaging crossover, \(\lambda \ge \tlam\). 
As one can see, the enhancement ratio is suppressed by increasing the coupling strength, and both results, Eq.~\eqref{eq:3-6} and Eq.~\eqref{eq:3-8}, converge to the mean-field theory.
This behavior can be explained by the competition between the coupling \(\lambda\) and the disorder \(h\).
As \(\lambda > h\), the coupling strength is dominant and the formation of the Cooper pairs is local, not affected by the realisation of the disorder potential.
On the other hand, when \(\lambda < h\), the potential takes the main role in defining which state and its time-reversal partner are to be coupled. 
This results in further enhancement in critical temperature due to the multifractality of the wavefunction and larger variance due to the sensitivity to disorder realization.

%\section{No self-averaging crossover in coupling strength}
\section{Breakdown of self-averaging and crossover in the coupling strength}
\label{sec:self-av}

We have seen in Fig.~\ref{fig:3-1} that the variance of the average critical temperature increases significantly for small enough couplings \(\lambda\).
In this section, we discuss the breakdown of the self-averaging of the critical temperature and quantify the crossover coupling strength \(\tlam\). 

In order to define the crossover coupling strength \(\tlam\), we use the equation~\eqref{eq:3-8} from which one extracts the critical temperature,
\begin{gather}
    \lambda W(T) \,{\boldsymbol\Delta} = {\boldsymbol\Delta}\,.
\end{gather}
It is an eigenproblem equation for the matrix \(W\) with the following matrix elements
\begin{gather}
    W_{nk}(T) = \frac{M_{nk}}{2\nu_0\epsilon_k}\,\tanh \left( \frac{\epsilon_k}{2T} \right)\,.
\end{gather}
Note that \(W\) depends explicitly on the disorder realization though the eigenvalues \(\epsilon_k\) and eigenstates of the Fibonacci chain appearing in \(M\)~\eqref{eq:3-2}.
It directly follows from the above equations, that for a given realization of the Fibonacci potential, the superconducting instability at some finite \(T\) exists only if the largest eigenvalue \(\Lambda(T=0)\) of \(W(T=0)\) is greater than \(1/\lambda\),
or equivalently \(\lambda\geq 1/\Lambda(T=0)\).
%we first find the weakest coupling strength \(\lambda_c\) for which a nontrivial solution (\(\Delta_n \neq 0\)) of Eq.~\eqref{eq:3-8} exists at zero temperature for a given disorder realization.
%Therefore there is a finite-\(T\) superconducting instability only if the smallest eigenvalue \(\Lambda(T)\) of \(W^{-1}(T)\) becomes smaller than the value of the superconducting coupling \(\lambda\ll 1\) (weak coupling limit) with decreasing  temperature.
%The crossing \(\Lambda(T_c)=\lambda\) defines the critical temperature, \(T=T_c\). 
%For the crossing to exist at any finite temperature, we must have \(\Lambda(T=0)\leq\lambda\).
%Therefore, superconductivity in a given realisation of the Fibonacci potential occurrs only if the attractive coupling strength \(\lambda\) exceeds some critical value, defined by \(\Lambda(T=0)\). 
%If the coupling \(\lambda\) is smaller than the eigenvalue \(\Lambda(T=0)\) for all disorder realizations, 
Based on this and the finite number of realizations of the Fibonacci potential for a given system size \(L\), we define
\begin{gather}
    \label{eq:4-1}
%    \tlam = \underset\mathrm{realizations}{\min}\Lambda(T=0) \\
    \tlam^{-1} = \min_{\{h_i\}}\Lambda(T=0)\,,
%    \lambda_e^{-1} = \max_{\{h_i\}}\Lambda(T=0)\,,   
\end{gather}
where the \(\min\) is taken over the realizations of the Fibonacci potential. 
The coupling \(\tlam\) corresponds to the appearance of the first disorder realization without a supercondicting phase. %\(T_c=0\), while \(\lambda_e\) marks the largest coupling where none of the realizations of the potential is superconducting.
%the system does not have a superconducting phase, for all Fibonacci realizations, and the critical temperature must be treated as zero. 
%For the superconducting couplings above the critical value, \(\lambda \leq {\tilde\lambda}\), at least some realisations will demonstrate superconductivity with a distribution of transition temperatures.
%\alexei{AA: do we have any statements/results for \(\lambda_e\)? e.g. is it finite in the thermodynamic limit or goes to zero?}
%\meng{Meng: There is no clearly tendency. This result is strongly related to the specific realization.}

\begin{figure}
    \centering
    \includegraphics[width=0.48\textwidth]{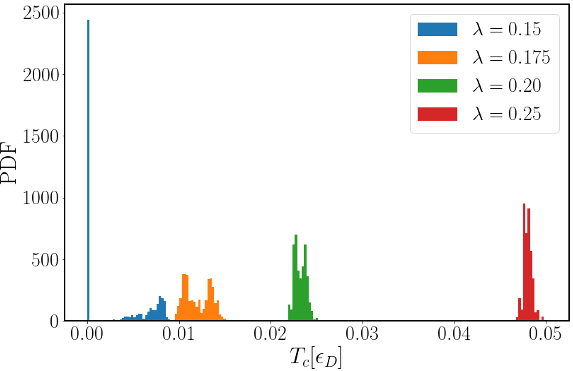}
    \caption{
        The PDF of the critical temperatures for disorder strength \(h=0.3\) and \(L=4000\) for different coupling strengths. 
        The crossover coupling strength is \(\tilde{\lambda}\approx 0.16\).
        }
    \label{fig:4-1}
\end{figure}

We now demonstrate that \(\tlam\) provides a proper definition of the crossover coupling, below which the self-averaging property of \(T_c\) is lost.
The naive argument is as follows: for \(\lambda < \tlam\) more and more disorder realizations stop having a superconducting phase, therefore increasing the sample to sample fluctuations, and making the average less well defined.
In Fig.~\ref{fig:4-1}, we show the probability density distributions (PDF) of critical temperatures for several values of \(\lambda\) with \(\tilde{\lambda} \approx 0.16\). 
We observe that as the coupling strength is decreased the average \(T_c\) becomes less representative.
For \(\lambda > \tilde{\lambda}\), the PDF has a bell shape and can be reasonably well approximated by a Gaussian, for instance for \(\lambda = 0.25\).
Closer to the crossover value \(\tilde{\lambda}\), the distributions (green and yellow) spread out, and several close peaks appear in the PDF.
For \(\lambda < \tlam\), the distribution continues to spread, acquires a visible tail for smaller \(T_c\) and the trivial case \(T_c = 0\) starts to accumulate (blue). 
As a consequence, the standard deviation of the critical temperature increase significantly.

\begin{figure}
    \centering
    \includegraphics[width=0.48\textwidth]{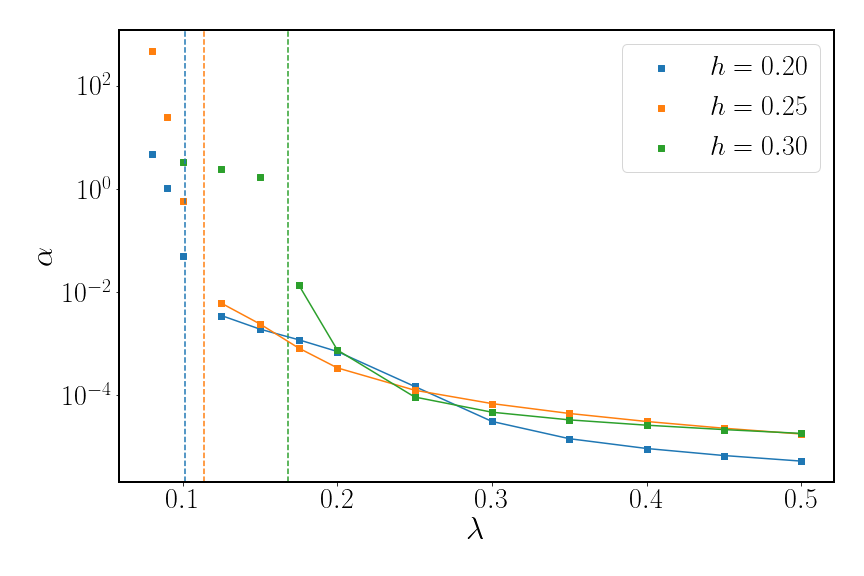}
    \caption{
%        Scatter points: 
        The self-averaging metric \(\alpha\) vs \(\lambda\) for different disorder strengths \(h\) at system size \(L=4000\). 
        The vertical dashed lines denote the position of the crossover \(\tlam\) for the respective disorder strengths.
        The points in the self-averaging regime, \(\lambda > \tlam\), are connected by the solid lines.
        \(\alpha\) becomes of order \(1\) for \(\lambda\lesssim\tlam\).
    }
    \label{fig:4-2}
\end{figure}

To further investigate the crossover coupling strength and the breakdown of self-averaging of the critical temperatures, we study the following metric of self-averaging:
\begin{gather}
    \label{eq:4-4}
    \alpha = \frac{\langle T_c^2 \rangle}{\langle T_c \rangle^2}-1,
    %\alpha = \frac{\sigma^2 \left(T_c\right)}{\langle T_c \rangle^2}.
\end{gather}
which quantifies the fluctuations around the average compared to the average itself: it is zero for perfectly self-averaging quantity (with \(\delta\)-function distribution).
The values of order \(1\) indicate that fluctuations around the average become comparable to the average itself, and the self-averaging property is lost.
In Fig.~\ref{fig:4-2}, we show the values of \(\alpha\) computed for the Fibonacci chain for different disorder strengths \(h\) at system size \(L=4000\).
The vertical dashed lines indicate the position of the crossover \(\tlam\) for several disorder strengths.
The solid lines connect the points for couplings above the crossover \(\tlam\).
%The solid curves are the least square fits to the self-averaging parameter \(\alpha\) truncated at the crossover \(\tlam\).
%\meng{The fitting function is chosen by experience as $\log_{10} f\left( x,a,b\right) =a\exp{-\frac{x}{b}-5}$ where $a$ and $b$ are the parameters to be determined. When $\lambda$ is closed to the crossover point, the calculation of \(T_c\) converges extremely slow by solving Eq.~\eqref{eq:3-8}. Thus directly calculating \(\alpha \left(\tlam\right)\) is not applicable. Instead, by the fitting functions, we estimate to self-averaging rate at crossover point as \(\alpha \left( \tlam \right)\sim 10^{-2}\), and their values gradually grow as we increase disorder strength.}
%\alexei{
%By the fitting functions, we estimate the self-averaging rate is at \(\alpha\left(\tilde{\lambda}\right) \sim 10^{-2}\), and their values gradually grow as we increase the disorder strength.
%AA: unclear
%}
We observe from Fig.~\ref{fig:4-2} that the two definitions of the crossover from self-averaging to no self-averagaing, \(\alpha\) and \(\tlam\), agree well.
The jump of the self-averaging parameter \(\alpha\) between the left and right side of \(\tilde{\lambda}\) from values of order \(10^{-2}\) to values of order \(1\), occurs precisely around the coupling \(\tlam\).
%which again shows the consistency between the two alternative definitions of the crossover in terms of the coupling strength.

\begin{figure}
    \centering
    \includegraphics[width=0.48\textwidth]{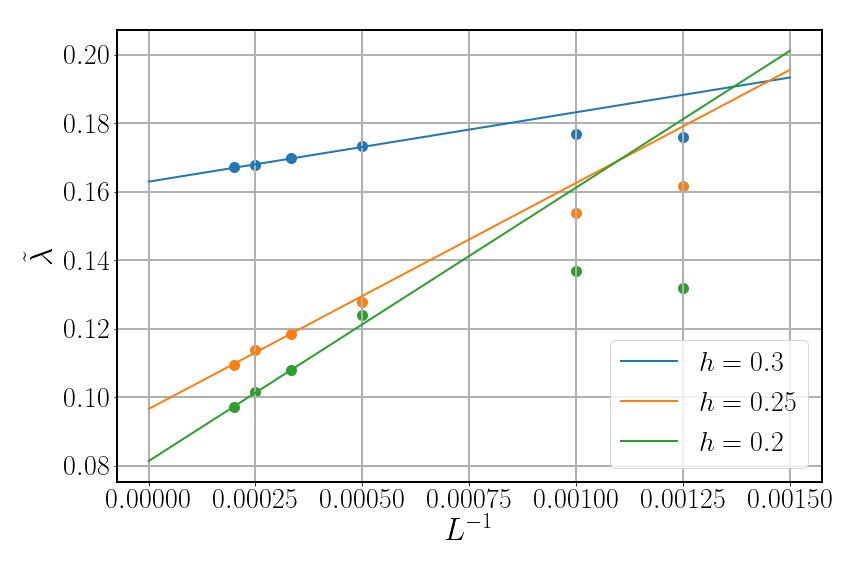}
    \caption{
        Finite-size scaling for the crossover coupling \(\tlam\) vs system size \(L\). 
        The fits are based on the data for the four largest sizes: \(L = 2000, 3000, 4000, 5000\).
    }
    \label{fig:4-3}
\end{figure}

Lastly we extract the thermodynamic limit of \(\tlam\) by extrapolation as shown in Fig.~\ref{fig:4-3}: 
extrapolation suggests finite values of \(\tlam\) in the thermodynamic limit.
Also the crossover \(\tlam\) decreases with decreasing the disorder strength.
This behavior can be anticipated as follows: for \(h=0\), the system is described by the BCS theory and \(T_c \sim \exp(-1/\lambda)\).
For a finite disorder, the non-zero values of \(\tlam\) indicate a transition from the superconducting phase to the insulator phase.
The observed dependence of \(\tlam\) with disorder strength naturally connects the two limits.

\section{Conclusions}

In this work, we considered an open \(1D\) chain with the Fibonacci potential \(h\) and calculated correlation of two single particle wavefunction for different disorder strengths \(h\).
We found a power-law behavior of the correlation function, which reflects the multifractal character of the eigenstates of the Fibonacci chain. 
%\alexei{
%Within the exact single-particle eigenstates approach, we calculated the critical temperature from two different routes: 
%AA: unclear!
%}
Using the single particle eigenstates, we used the mean-field theory to compute the critical temperature of the superconducting transition following two different procedures:
\begin{itemize}
    \item averaging Eq.~\eqref{eq:3-8} first, then solving it for \(T_c\), e.g. by averaging spatial correlation function \(C(\omega)\) and estimating the multifractal related parameters \(\gamma\) and \(E_0\), we analytically calculated the critical temperature via Eq.~\eqref{eq:3-6} assuming self-averaging of all characteristic variables (as explained in the above text);
    \item first solving Eq.~\eqref{eq:3-8}, then averaging, e.g. by solving Eq.~\eqref{eq:3-8} numericaly for the critical temperature for a fixed realization of the Fibonacci potential, and analyzing the statistics -- PDF, mean value and variance -- of the ensemble of critical temperatures. 
\end{itemize}
We found a clear discrepancy between the results obtained with these two methods, which we attribute to neglecting correlations present in Eq.~\eqref{eq:3-8} between \(T_c\) and the single-particle eigenfunctions in the kernel \(M\), and eigenvalues \(\epsilon_m\).
%which we relate to the non-self-averaging property of the critical temperature and its correlation with the single-particle eigenfunctions. 
Our exact numerical approach clearly demonstrates the enhancement of the critical temperature in comparison to other approaches relying on neglecting the correlations in the equation~\eqref{eq:3-8} for the critical temperature.
%assuming self-averaging property of the critical temperature.

We observe that for strong enough couplings, critical temperature is self-averaging, however that breaks for weaker couplings.
We introduced the quantity \(\tlam\) to quantify the breakdown of the self-averaging property of the critical temperature.
When \(\lambda > \tlam\), the self-averaging is well preserved and the distribution of the critical temperature can be approximated by a Gaussian. 
On the other hand, when \(\lambda \leq \tlam\), the standard deviation start to spread significantly, indicating that the solution for the critical temperature becomes extremely sensitive to the disorder realization.

\begin{acknowledgements}
    T\v{C}, AA acknowledge the financial support from the Institute for Basic Science (IBS) in the Republic of Korea through the project IBS-R024-D1. 
    IVY gratefully acknowledges support from the Leverhulme Trust under the grant RPG-2019-317.
    While preparing this work we became aware of a closely related work, Ref.~\onlinecite{oliveira2023incommensurability}.
\end{acknowledgements}

%\appendix

\bibliography{general,mbl}
% local.bib
% mbl
% general

\end{document}